\documentstyle [12pt]{article}
\def\be{\begin{equation}}
\def\ee{\end{equation}}
\def\bea{\begin{eqnarray}}
\def\eea{\end{eqnarray}}

\begin{document}

\begin{titlepage}

\title{$\sigma_1$ and $\sigma_2$ automorphism of the systems of equations of integrable hierarchies.\\Discrete and Backlund transformations}

%\Explicit formula for multisoliton solutions

\author{A.N. Leznov\\Universidad Autonoma del Estado de Morelos,\\
CIICAp,Cuernavaca, Mexico}

\maketitle

\begin{abstract}

It is shown that there exists two inner authomorpism which lead to different form  of the sistems equations of integrable hierarchy. We present discrete and Backlund transformation connected with such systems and a general formula for multi-soliton solutions based on these symmetries.

\end{abstract}
\end{titlepage}

\section{Introduction}

In a series of papers by the author starting in 80's  of the  previous century \cite{i} a simple method for constructing integrable systems  together with their
soliton-like solutions was proposed. This method requires only two calculational steps -- solving a system of linear algebraic equations and performing a Gaussian decomposition of a polynomial into a product determined by its roots. This lead to a nonlinear symmetry of integrable system termed a discrete substitution or an integrable mapping. A discrete substitution is a nonlinear transformation (with no additional parameters) that, given an arbitrary solution of an integrable system, produces a new solution according to certain rules.

All systems of equations invariant with respect to such a mapping are united into a hierarchy of integrable systems that bears  the name of the corresponding discrete substitution. This substitution is a canonical transformation \cite{7},\cite{8} which explains successful applications of  methods of canonical transformation theory to integrable systems \cite{FT}.

Let us assume that an integrable system has some inner automorphism $\sigma$. This
implies that there is a solution that is also invariant under $\sigma$. Usually, such  solutions are interesting for applications. In general, a discrete transformation will not commute with $\sigma$. However, it is possible starting from a non-invariant solution to produce a ``good'', invariant, solution after several applications of the discrete transformation (implementation of this scheme is outlined Refs.~\cite{10} and \cite{11}).

On the other hand, it is well known that Backlund transformations contain additional parameters and therefore allow to increase the number of parameters in the solution after their application. Because the method of discrete substitution allowed to obtain n-soliton solutions directly and  explicitly, the author had no interest in Backlund transformations from point of view obtaining soliton solutions. But Backlund transformation have more wide sence and we would like to close this loophole in the present paper.  

The goal of the present paper is to show how to construct Backlund transformations for integrable systems using methods of \cite{10},\cite{11}. We  restrict ourselves  with the well-known example of the nonlinear Schrodinger equation to ease the demonstration of the main idea and methods needed for solution of this problem. The extension to the general case is obvious and will be clear after considering this single example.

\section{Discrete transformation}

All considerations of the present paper are given for the simplest  example of the nonlinear Schrodinger equations. However,  it will become clear from our arguments, detailed calculations below, and references that our method is applicable to all integrable systems with soliton-like solutions.

The nonlinear Schrodinger equation can be represented in two forms:
the classical one for one complex-valued unknown function $\psi(x,t)$
\be 
-2i\psi_t+\psi_{xx}+2(\psi\bar \psi)\psi=0\label{CNL} 
\ee 
and as a wider system of two equations for two unknown complex-valued functions functions $u,v$ \be 
-i2 u_t+u_{xx}+2(uv)u=0, \quad i2v_t+v_{xx}+2(uv)v=0 \label{SCNL} 
\ee
(The factor of 2 multiplying time derivatives can be of course eliminated by redefining the time variable.) The last system is obviously invariant under an inner automorphism $\sigma_1$: $u\to
\bar v,v\to \bar u$ and thus it has solutions invariant with respect to this change of variables. Such solutions satisfy $u=\bar v=\psi$ and are therefore also solutions of (\ref{CNL}).

The system (\ref{SCNL}) is invariant with respect to the following nonlinear invertible transformation 
\be 
U={1\over v},\quad V=v(uv-(\ln v)_{xx}),\quad v={1\over U},\quad u=U(UV-(\ln U)_{xx})\label{DM} \ee 
which is referred to as a discrete transformation, discrete substitution or an integrable mapping.

Transformation (\ref{DM}) takes a given solution $(u,v)$ of the system (\ref{SCNL}) to a new one $(U,V)$. But this transformation is not invariant with respect to $\sigma_1$. Thus, if the initial solution has a property $(u=\bar v)$, the new solution $(U,V)$ will not have this property. Now we make the following very important observation. If we apply the direct discrete transformation $m$ times to a solution of (\ref{CNL}) $(u=\bar v)$ with the result
$U^{+m},V^{+m}$ and apply the inverse transformation to the same solution $m$ times with the result $u^{-m},v^{-m}$,  the resulting new solutions are related as follows: $U^{+m}=\bar
u^{-m},V^{+m}=\bar v^{-m}$. Thus, if we start from an obvious linear solution of the system (\ref{SCNL}) $u_0=0, 2i{v_0}_t+{v_0}_{xx}=0$ and after $2m$ steps of direct discrete transformation obtain a solution of $2i{u_{2m}}_t+{u_{2m}}_{xx}=0,v_{2m}=0$ with
$u_{2m}=\bar v_0$, we are guaranteed that at the $m$th step we will obtain a (m-soliton) solution of (\ref{CNL}) -- traditional nonlinear Schrodinger equation. This approach together with the corresponding mathematical formalism is described in detail in \cite{10},\cite{11}.

\section{Backlund transformation}

Now we would like to find  not the symmetry of the enlarged system
(\ref{SCNL}) but independently the symmetry of (\ref{CNL}). This symmetry transformation can contain additional numerical parameters and after each its application we increase the number of parameters in the solution. This is exactly  Backlund's original idea. However, one has to keep in mind that, applied to the general solution of the equation, this transformation cannot add any new independent parameters, but can only change the initial functions on which the
general solution depends.

Consider a solution of the classical nonlinear Shrodinger equation $\psi=u=\bar v$. This equation may be written in Lax pair form
\be
g_xg^{-1}=i\pmatrix{ \lambda, & u \cr
                      v & -\lambda \cr}\quad
g_tg^{-1}=i\pmatrix{ \lambda^2 -{uv\over 2}, & \lambda u-i{u_x\over 2} \cr
                      \lambda v+i{v_x\over 2} & -\lambda^2+{uv\over 2} \cr}\ \label{LA}
\ee 
Now let us use the formalism of \cite{11} and try to find a new
solution of the classical nonlinear Schrodinger equation in the form
\be 
G=\pmatrix{\lambda+A & B \cr
               C & \lambda +D \cr}g\equiv Pg \label{IMP}
\ee 
where functions $A,B,C,D$ are determined by imposing the condition that matrix $G$ has linear dependence between its columns (and rows) with constant coefficients $(c^i_1,c^i_2)$ at two points of the complex $\lambda$ plane, $\lambda_{1,2}$. In the Appendix we show that this  condition is not independent but follows directly
from (\ref{IMP}).

From this condition we immediately obtain $Det P=(\lambda-\lambda_1)(\lambda-\lambda_2)$ and a linear system of equations for relating functions $A,B,C,D$ to matrix elements of $g$, parameters $\lambda_i$, and vectors $c^i$
\be
(\lambda_1+A)(gc)^1_1+B(gc)^1_2=0,\quad
(\lambda_2+A)(gc)^2_1+B(gc)^2_2=0\label{LS1}
\ee 
\be
C(gc)^1_1+(\lambda_1+D)(gc)^1_2=0,\quad
C(gc)^2_1+(\lambda_1+D)(gc)^2_2=0\label{LS2} 
\ee 
In particular, we have 
\be 
(\lambda_1-\lambda_2)+B({(gc)^1_2\over
(gc)^1_1}-{(gc)^2_2\over (gc)^2_1})=0,\quad
(\lambda_1-\lambda_2)+C({(gc)^1_1\over (gc)^1_2}-{(gc)^2_1\over (gc)^2_2})=0\label{ESP}
\ee 
Now let us obtain a new L-A pair. For the matrix element $(G_xG^{-1})_{11}$ we
have 
\be 
(G_xG^{-1})_{11}={det \pmatrix{A_x+(\lambda+A)i\lambda+iBv & B_x+(\lambda+A)iu-Bi\lambda \cr
                                    C & \lambda  +D \cr}\over det P}=i\lambda
\ee
where we used the fact that the determinants in the numerator and denominator have zeros at the same points. Similarly, we obtain
\be
(G_xG^{-1})_{12}={det \pmatrix{ \lambda +A &  B \cr
A_x+(\lambda+A)i\lambda+Biv & B_x +(\lambda+A)iu-Bi\lambda \cr}\over det P}=i(u-2B)
\ee
and finally
\be
G_xG^{-1}=i\pmatrix{ \lambda, & u -2B\cr
                         v+2C & -\lambda \cr}\label{!!}
\ee

If we did not demand the condition of complex conjugation $u=v^*$ to hold,  this would exactly be a Backlund transformation for the enlarged nonlinear system (\ref{SCNL}).

But if we start from a solution of the nonlinear Schredinger equation (\ref{CNL}) $u=v^*$ and would like to obtain a solution of the same equation, we have to require, according to the last expression, that $\bar C=-B$.

To this end, it is necessary to use known explicit expressions (\ref{ESP}). From L-A representation (\ref{LA}) it follows that $g$ can be considered as a unitary matrix $gg^H=1$
under assumption that $\lambda$ is real.

This means that $g^{-1}=g^H$ or in a matrix form (in what follows $\bar f\equiv f^*)$
$$
g(\lambda)_{22}=g^*(\lambda)_{11},\quad g(\lambda)_{12}=-g^*(\lambda)_{21},
$$
Taking into account that the matrix elements of $g$ are analytic functions of
$\lambda$ (as solutions of a differential equation with coefficients analytic in $\lambda$), we conclude that
$$
(g(\lambda)_{22})^*=g(\lambda^*)_{11},\quad (g(\lambda)_{11})^*= g(\lambda^*)_{22},
$$
$$
(g(\lambda)_{21})^*=-g(\lambda^*)_{12},\quad (g(\lambda)_{12})^*=-g(\lambda^*)_{21}
$$
Now we can evaluate $\bar C$. We have
\be
{(gc)^1_1\over (gc)^1_2}={g(\lambda_1)_{11}c^1_1+g(\lambda_1)_{12}c^1_2\over
g(\lambda_1)_{21}c^1_1+g(\lambda_1)_{22}c^1_2}={g(\lambda_1)_{11}+g(\lambda_1)_{12}\alpha_1\over g(\lambda_1)_{21}+g(\lambda_1)_{22}\alpha_1}\label{MIDL}
\ee
where $\alpha_1={c^1_2\over c^1_1}$. According to the above formulae for complex conjugation,
we have
$$
(F(\lambda_1,\alpha_1)^*\equiv({(gc)^1_1\over (gc)^1_2})^*={g(\lambda^*_1)_{22}-g(\lambda^*_1)_{21} \alpha^*_1\over -g(\lambda^*_1)_{12}+g(\lambda^*_1)_{11}\alpha^*_1}=
-{1\over F(\lambda_1^*,-{1\over \alpha^*_1})}
$$
For further manipulations let us rewrite (\ref{ESP}) in terms of $F_1,F_2$ introduced above
\be
(\lambda_1-\lambda_2)+C(F(\lambda_1,\alpha_1)-F(\lambda_2,\alpha_2))=0,\quad
(\lambda_1-\lambda_2)+B({1\over F(\lambda_1,\alpha_1)}-{1\over F(\lambda_2,\alpha_2)})=0\label{MIDL1}
\ee
The condition $B=-C^*$ together with the conjugation properties of functions $F$ lead to
\be
\lambda_1=\lambda^*_2,\quad \alpha^*_1=-{1\over \alpha_2}\label{BEK!}
\ee
Thus if for some solution of (\ref{CNL}) the corresponding element $g$ is known new solution of the same equation differente from the initial on the pair of complex parameters
can be constructed by the rules of present section. (But not forget about the comments on this subgect in the beggining of this section). 

\section{Multisoliton solutions}

Now let us apply $n$ times the transformation of the previous section to a certain solution of the nonlinear Shrodinger equation (\ref{CNL}). Each transformation is defined by two complex parameters $\lambda_i,\alpha_i$. After n applications, we have for the corresponding matrix $G_n$
$$
G_n=\pmatrix{\lambda+A & B \cr
                    C & \lambda +D \cr}G_{n-1}=
\pmatrix{\tilde P^n_{11} & P^{n-1}_{12} \cr 
        P^{n-1}_{21} & \tilde P^{n}_{22} \cr}g_0
$$
Here the notation $\tilde P$ means that the coefficient at the highest degree of the corresponding polynomial is equal to one. $G_n$ has zero vectors at $2n$ points of the $\lambda_i,\lambda_i^*$ plane with the coefficients of proportionalities $\alpha_i,-{1\over \alpha_i^*}$. For this reason all coefficients of polynomials of the element $G_n$
can be obtained from the linear system of equations
$$
\lambda_i^n+\sum_{k=0}^{n-1} P^k_{11}\lambda_i^k +\sum_{k=0}^{n-1} F(\lambda_i,\alpha_i)P^k_{12} \lambda_i^k=0
$$
$$
\sum_{k=0}^{n-1} F^{-1}(\lambda_i,\alpha_i)P^k_{21} \lambda_i^k +\lambda_i^n+\sum_{k=0}^{n-1} P^k_{22}\lambda_i^k=0
$$
where $P^k_{ij}$ are the coefficients at $\lambda^k$ in the corresponding polynomial.
In connection with  Kramers rules coefficients interesting for further considerations are
\be
P^{n-1}_{12}=-{det_{2n} (\lambda_i^{n-1},..,\lambda_i,1;\lambda_i^n;F_i\lambda_i^{n-2},...F_i\lambda_i,F_i)\over det_{2n}(\lambda_i^{n-1},..,\lambda_i,1;F_i\lambda_i^{n-1},...F_i\lambda_i,F_i)}\label{FIN}
\ee
\be
P^{n-1}_{21}=-{det_{2n}(\lambda_i^n;F^{-1}_i\lambda_i^{n-2},...F^{-1}_i\lambda_i,F^{-1}_i;\lambda_i^{n-1},..,\lambda_i,1)\over det_{2n}(F^{-1}_i\lambda_i^{n-1},...F^{-1}_i\lambda_i,F^{-1}_i;\lambda_i^{n-1},..,\lambda_i,1)}\label{FINI}
\ee
In the last formula we symbolically wrote the structure of each of $2n$ lines of the corresponding determinant matrices. In the case of a diagonal initial matrix $g_0=\exp i(\lambda x+\lambda^2 t) h$ formula (\ref{FIN}) was presented in \cite{10} without any connection to
the Backlund transformation of the present paper.

Using absolutely the same technique as in the previous section we obtain
\be
(G_n)_xG_n^{-1}=i\pmatrix{ \lambda, & u -2P^{n-1}_{12}\cr
                    v+2P^{n-1}_{21} & -\lambda \cr}\label{!!!}
\ee
Thus, after n steps of Backlund transformation each of which is defined by parameters
$(\lambda_i,\lambda_i^*, \alpha_i, -{1\over \alpha_i^*})$, we get a new solution of 
the nonlinear Shrodinger equation
$$
U=u-2P^{n-1}_{12},\quad V=v+2P^{n-1}_{21}
$$
From the explicit expressions (\ref{FIN}) and (\ref{FINI}) it follows that all Backlund transformations are  commutative ( functions $P^{n-1}_{12},P^{n-1}_{21}$ are symmetrical to permutation of all pairs $\lambda_i,\alpha_i$).

If we choose the initial solution in a "zero" form $u=v^*=0,\quad g=\exp i2(\lambda x+\lambda^2 t) h$, we obtain an n-soliton solution in an explicit form. This solution of course coincides with the one obtained previously in \cite{10},\cite{11} using the method of discrete transformation and repeated in the previous section.

The result of \cite{10} corresponds to decomposition of the determinant of $2n$th order into a sum of the products of the corresponding minors  $n$th order.

\section{Backlund transformation in its original form}

As it follows from the material of the previous section, Backlund transformation, in contrast to the discrete one,  is not local. Indeed to obtain a new solution $U=V^*$
from a solution $u=v^*$, we have to resolve the L-A system (which is a  system of linear differential equations), obtain the element $g$ and use its matrix elements  to construct a new solution.

This fact is related to the original Backlund's result, who  found relations between the derivatives of the new and old solutions. We would like to show now, how such relations can be obtained from the formalism of the previous section.

Using the definition of $F(\lambda_i,\alpha_i)$ and taking into account equations of L-A pair (\ref{LA}), we obtain
$$
F^s_x\equiv (F(\lambda_s,\alpha)_s)_x=i(v-2\lambda_s F^s-u(F^s)^2),
$$
$$
F^s_t=i((\lambda_s v+i{v_x\over 2})-2(\lambda_s^2-{uv\over 2})F^s-(\lambda_su-i{u_x\over 2})(F^s)^2)
$$
Calculating $F^1,F^2$ via $2B=u-U,2c=V-v$ from (\ref{MIDL1}) and substituting result into the system of derivatives above (with respect to $x$ or $t$ arguments) we come to a system of two ordinary differential equatinos of the first order connected  $u,v$ and $U,V$ and containing parameters $\lambda_1=\lambda^*_2$ in explicit form. Two additional parameters $\alpha_1, \alpha_2=-{1\over \alpha^*}$ have to arised in process of integration of the last system in which $u=v^*$ considered as known and $U=V^*$ as unknown functions or visa versa.

\section{Second automorphism $\sigma_2$ of nonlinear Schrodinger system}

Let us perform a change of variables $v=e^{i\theta}, u=e^{-i\theta}(R-{i\theta_{xx}\over 2})$ in (\ref{SCNL}), where $R,\theta$ two new unknown complex function.
The sence of this substitution will be clarified something later.
Instead of (\ref{SCNL}) we have
\be
2\theta_t+(\theta_x)^2=2R,\quad 2R_t-{1\over 2}\theta_{xxxx}-2(\theta_x R)_x=0\label{SCHL1}
\ee
Of course the last system is invariant with obvious exchange $R\to R^*,\theta\to
\theta^*$, which we call second inner automorphism $\sigma_2$ of nonlinear Shrodinger system.

The system (\ref{SCHL1}) after excluding $R$ is equivalent to single equation
\be
\theta_{tt}-{1\over 4}\theta_{xxxx}+({3\over2}\theta_x^2+\theta_t) \theta_{xx}=0
\label{s2}
\ee
The Author cannot say anything about the system (\ref{SCHL1}) or equation (\ref{s2}) because as it seems he ever encountered them before in such a form in the literature.

\subsection{Method of discrete substitution}

The nature of $\sigma_2$ authormorphism consists in the fact that if discrete transformation (\ref{DM}) is iterrupted on $2n+1$ step (but not on $2n$ one as it was in the case of $\sigma_1$), it has as it conclusion (\ref{s2}). Indeed in this case in the middle of the latice arises two solutions ($u_n={D_{n-1}\over D_n}, v_n={D_{n+1}\over D_n}$) and ($u_{n+1}={D_n\over D_{n+1}}, v_{n+1}={D_{n+2}\over D_{n+1}}$), which are connected by condition $v_n^*=u_{n+1}={1\over v_n},\quad u_n^*=v_{n+1}=v_n(u_nv_n+(\ln v_n)_{xx})$, which lead to (\ref{SCHL1}) and (\ref{s2}).

Using  the technique of dicrete transformation n-soliton solution of the equation (\ref{s2})and system (\ref{SCHL1}) may be represented in the terms of the following notations
$$
F=\sum_{k=1}^{2n+1} c_k e^{i2L_k},\quad f=\sum_{k=1}^{2n+1} {1\over c_k} e^{-iL_k}{1\over \prod'
(\lambda_k-\lambda_j)^2}
$$
where $L_i=\lambda_i^2t+\lambda_ix$ and $\lambda_i,c_i$ -$(2n+1)$ numerical parameters connected via equation $F^*=f$.
From the last condition it follows the limitation on parameters of the problem: $2s+1$ parameters $\lambda_{\beta}^*=\lambda_{\beta}$ are real one, the remaining
$2(n-s)$ are in complex congugated pairs $\lambda_A^*=\lambda_B$ ($1 \leq A,B\leq (n-s)$). In all cases $c_ic_i^*=\prod'(\lambda_i-\lambda_j)^2$. This number under all chossing of $\lambda$ parameters above is a positive number. Finally
$$
\theta=i\ln {D_n\over D_{n+1}},\quad R=(\ln(D_nD_n^*))_{xx}
$$
where $D_s$ is determinant of the s- order of the matrix (typical for discrete transformation calcules):

$$
\pmatrix{ F & F_x & F_{xx} &....\cr
  F_x & F_{xx} & F_{xxx} &...\cr
F_{xx} & F_{xxx} & F_{xxxx} &...\cr}
$$

\subsection{Method of the Backlund transformation}

It is necessary to repeat word by word all calculations of the section 3
up to (\ref{!!}), keeping in mind that in all formulae connected $g$ and ($u,v$) now
$v=e^{i\theta},u=e^{-i\theta}(R-{i\over 2}\theta_{xx})$.

"One" soliton solution, the simplest solution of (\ref{SCHL1}) is the following one
$(\theta=-2(\lambda_0^2t+\lambda_0x)\equiv {-2L_0}, R=0)$. It is obvious that correspoding group element $g$ belongs to the group of lawer triangular matrices and may be represented in the following form:
$$
g(x,t:\lambda)=e^{\alpha X_-}e^{\tau H}
$$
Taking into account equations of L-A pair (\ref{LA}) we obtain in a consequence
$$
g_xg^{-1}=(\alpha_x+2\tau_x\alpha)X_-+\tau_xH=i(\lambda H+e^{-2iL_0}X_-)
$$
$$
g_tg^{-1}=(\alpha_t+2\tau_t\alpha)X_-+\tau_tH=i(\lambda^2 H+e^{-2iL_0}(\lambda+\lambda_0)X_-)
$$
Solition of the last equations are trivial with the finally result
$$
\alpha={1\over 2}{e^{-i2L_0}\over \lambda-\lambda_0},\tau=i(\lambda^2t+\lambda x)
$$

In the general case let us represent element $g$ in ussual form of $S(2,C)$ group
\be
g=e^{\alpha X_+}e^{\tau H} e^{\beta X_-}=
\pmatrix{e^{\tau}+\alpha \beta e^{-\tau} & \alpha e^{-\tau} \cr
\beta e^{-\tau} & e^{-\tau} \cr}\label{PAR}
\ee
\be
g'g^{-1}=(\alpha'-2\tau'\alpha-\beta'e^{-2\tau}) X_++(\tau'+\beta'\alpha e^{-2\tau})H+
\beta'e^{-2\tau}X_-\label{PAR1}
\ee
Taking into acount equations of L-A pair (\ref{LA}) we obtain
$$
i\lambda=(\tau_x+\beta_x\alpha e^{-2\tau}),\quad iv=\beta_x e^{-2\tau}
$$
From the last relation we conclude that $\beta'e^{-2\tau}$ doesn't depend from $\lambda$ and
$$
\alpha={i\lambda-\tau_x \over \beta_x e^{-2\tau}}
$$
Now let us rewrite equation defining $C$ (\ref{ESP}) substituting in it (\ref{PAR}) ( to have no mixing we change parameters $\alpha_{1,2}$ in (\ref{MIDL}) on $\nu_{1,2}$)
\be
{1\over C}+{1\over \lambda_1-\lambda_2}({e^{2\tau_1}\over \beta_1+\nu_1}+\alpha_1-
{e^{2\tau_2}\over \beta_2+\nu_2}-\alpha_2)=0\label{!!!}
\ee
The new solution $V=v+2C$ (result of transformation $V=v+2C$) (\ref{!!}) and condition of its inariantnes with respect to $\sigma_2$ $VV^*=1$ may be rewritten as ($vv^*=1 !$)
\be
{v\over C}+{v^*\over C^*}=-2\label{CC}
\ee
or using  (\ref{!!!}) we have in a consequence $(iv=\beta_x e^{-2\tau})$
$$
{v\over C}+{\beta_xe^{-2\tau}\over i(\lambda_1-\lambda_2)}({e^{2\tau_1}\over \beta_1+\nu_1}+\alpha_1-{e^{2\tau_2}\over \beta_2+\nu_2}-\alpha_2)=0
$$
Exept of this from (\ref{CC}) ($v^*={1\over v}$) we obtain also
$$
v=C(-1+i\sqrt {{1\over CC^*}-1}),\quad V=C(1+i\sqrt {{1\over CC^*}-1})
$$

Keeping in mind all obtain above expresions and comments we rewrite the last equality in the form
$$
{v\over C}+1+{1\over i(\lambda_1-\lambda_2)}[{\beta^1_x\over \beta^1+\nu_1)}-\tau^1_x-
{\beta^2_x\over \beta^2+\nu_2)}-\tau^2_x]=0,
$$
$$
{v\over C}+1+{1\over i(\lambda_1-\lambda_2)}
[\ln {(\beta^1+\nu_1)e^{-\tau_1}\over (\beta ^2+\nu_2)e^{-\tau_2}}]_x=0
$$
Summarizing with complex congugation and not forgetting about (\ref{CC})
we come to eqution
\be
{1\over i(\lambda_1-\lambda_2)}[\ln {(\beta^1+\nu_1)e^{-\tau_1}\over (\beta^2+\nu_2)e^{-\tau_2}}]_x-{1\over i(\lambda_1^*-\lambda_2^*)}[\ln {(\beta^1+\nu_1)e^{-\tau_1}\over (\beta^2+\nu_2)e^{-\tau_2}}]^*_x=0\label{VIC}
\ee
To find relation connected 4 up to now arbitrary numerical parameters $\lambda_{1,2},\nu_{1,2}$ it is necessary to take into account second condition of congugation dectated by authomrphism $\sigma_2$, which allow to find connection $\beta$ and $\beta^*$ functions.

Comparing (\ref{PAR1}) and L-A pair representation (\ref{LA}), we obtain
$$
(\alpha'-2\tau'\alpha-\beta'e^{-2\tau})=i u\equiv i {1\over v} uv=i{1\over v}(r-{1\over 2}(\ln v)_{xx})
$$
Substituting all above obtained relations we have
$$
-r+{i\over 2}\theta_{xx}=\lambda^2-\tau_{x,x}+(\tau_x)^2 +(\lambda+i\tau_x)\theta_x
$$
In a L-A pair formalism $\lambda$ have to be considered as a real parameter. Thus
imiginary part of last equation has the form
$$
\theta_{xx}={1\over i}(\tau^*-\tau)_{xx}+(\tau^*+\tau)_x\theta_x+(\tau_x)^2-(\tau^*_x)^2
$$
The last equation has obvious first integral
$$
\theta_x-{1\over i}(\tau^*-\tau)_x=c e^{(\tau^*+\tau)}
$$
Reminding that $ie^{\theta}=\beta_x e^{-2\tau}$ and substituting into the last
quation we transform it to the form
$$
{1\over i}(\ln \beta_x e^{-(\tau^*+\tau)})_x=c_1 e^{(\tau^*+\tau)}\quad
{1\over i}(\beta_x e^{-(\tau^*+\tau)})_x=c_1 \beta_x
$$
Second integration leads to final result
\be
{1\over i}(\beta_x e^{-(\tau^*+\tau)})=e^{i\theta}e^{-(\tau^*-\tau)}=c_1\beta+c_0\label{EXTRA}
\ee
we pay attention of the reader that in (\ref{EXTRA})parameters $c_1,c_0$ not depend from $(x,t)$ variables, but only from real parameter $\lambda$. Inverting (\ref{EXTRA}) 
we obtain $\beta$ in form in which it became clear its structure with respect to complex conjuguation
$$
\beta+\nu={1\over c_1}e^{i\theta}e^{-(\tau^*-\tau)}+\nu-{c_0\over c_1}
$$
Putting in the last expression indexes $k=1,2$ for $\lambda$ we come to elements introducing in (\ref{VIC})
$$
\beta_k+\nu_k={1\over c_1^k}e^{i\theta}e^{-(\tau^*_k-\tau_k)}+\tilde \nu_k
$$
where $\tilde \nu_k=\nu_k-{c_0^k\over c_1^k}$
$$
\beta_k^*+\nu^*_k={1\over (c_1^k)^*}e^{-i\theta}e^{(\tau^*_k-\tau_k)}+\tilde \nu_k^*
$$
After substitution the last expressions into (\ref{VIC}) and simplest manipulations we come to two possibilities. First one $\lambda_1=\lambda_1^*,\lambda_2=\lambda_2^*,\tilde \nu_1\tilde \nu_1^*={1\over c_1^1(c_1^1)^*},\tilde \nu_2\tilde \nu_2^*={1\over c_1^2(c_1^2)^*}$ and the second one $\lambda_1=\lambda_2^*,\tilde \nu_1\tilde \nu_2^*={1\over c_1^1(c_1^2)^*}$.
Of course this result in the case of multisoliton solutions coinsedes with obtained above by the method of discrete substitution.

\section{Outlook}

First, we would like to explain why, in spite of our comments in the Introduction,
Backlund transformations are interesting objects for investigation.
Backlund transformations, as well as  discrete ones, are canonical transformations. This means that under these transformations densities of conserved quantities change by complete derivatives. In the case we considered above this applies in particular to the energy density. Therefore, if the energy density of the initial solution is $e$,
after the application of the Backlund transformation it will be $e_B=e+\Delta_x$.
The term $\Delta$ contains all parameters $\lambda_1=\lambda^*_2,\alpha_1=-{1\over \alpha^*}$,
which define the Backlund transformation. For obtaining an explicit expression for $\Delta$
an additional (not a very cumbersome) calculation, using the technique of \cite{10},\cite{11}, is necessary. If $e$ itself has been obtained from some other solution and so on,  we will have a final expression
$$
e_n=\sum \Delta^i_x
$$
Each $\Delta^i$ contains only parameters of  previous transformations.
Because the total energy for an $n$-soliton solution is equal to n, it is very natural to assume that all terms of the last sum have the same behavior at infinity and contribute 1 to the total energy after integration.

It is completely obvious that $\sigma_1$ and $\sigma_2$  exist for all integrable systems possessing soliton solutions and a discrete transformation. Indeed, as it was explained above, this is related only to the parity of the step, odd (2n) or even (2n+1), on which the discrete transformation applied to the initial solution is interrupted.

Without any doubts all results of this paper also apply to supersymmetric integrable hierarchies \cite{LS} and multicomponent matrix-type hierarchies \cite{LY},\cite{LYY}.

What are the systems connected to $\sigma_2$ is completely unknown to the author at this moment.

\section{Acknowledgements}

Author indepeted to E.A.Yuzbajan for discussion of the results and big help in the process
of preparetion the manuscript for publication.

\section{APPENDIX}

In this Appendix we would like to show that 

Let us rewrite (\ref{LA}) and (\ref{IMP}) in equivalent form
$$
\pmatrix{A_x & B_x \cr
         C_x & D_x \cr}+\pmatrix{\lambda+A & B \cr
               C & \lambda +D \cr}i\pmatrix{ \lambda, & u \cr
                      v & -\lambda \cr}=i\pmatrix{ \lambda, & U \cr
                      V & -\lambda \cr}\pmatrix{\lambda+A & B \cr
                                                 C & \lambda +D \cr}
$$
$$
\pmatrix{A_t & B_t \cr
         C_t & D_t \cr}+\pmatrix{\lambda+A & B \cr
               C & \lambda +D \cr}i\pmatrix{ \lambda^2 -{uv\over 2}, & \lambda u-i{u_x\over 2} \cr
                      \lambda v+i{v_x\over 2} & -\lambda^2+{uv\over 2} \cr}=
$$
$$
i\pmatrix{\lambda^2 -{UV\over 2}, & \lambda U-i{U_x\over 2} \cr
 \lambda V+i{V_x\over 2} & -\lambda^2+{UV\over 2} \cr}
\pmatrix{\lambda+A & B \cr
            C & \lambda +D \cr}
$$
From which we have the following system of equations
$$
u-U=2B,\quad v-V=-2C,\quad iA_x=Bv-CU,\quad iD_x=Cu-BV,\quad iB_x=Au-DU,\quad
iC_x=Dv-AV
$$
and the same kind of the equations with respect to $t$ differetiation.
As result we have $A+D,AD-BC$ doesn't depend from $x,t$ arguments and thus representable in the form $A+D=\lambda_1+\lambda_2,AD-BC=\lambda_1\lambda_2$.
Besides this functions $C,D$ satisfy the following system of equations

$$
-iC_t-{1\over 2}( C_{xx}+4ikC_x+4k^2C)-C(BC)+{(C_x)^2B\over 4(\epsilon -CB)}=0
$$
$$
i B_t -{1\over 2}(B_{xx}-4ikB_x+4k^2B)-B(BC)+{(B_x)^2C\over 4(\epsilon -CB)}=0
$$
where $k=\lambda_1+\lambda_2,\epsilon=({\lambda_1-\lambda_2)\over 2})^2$.

The last system of course is equivalent to nonlinear Schrodinger equation. More other solution of (\ref{SCNL}) is connected with the solution of the above system via relations
$$
v=-C-{iC_x+kC\over 2\sqrt {\epsilon-BC}},\quad u=B+{iB_x-kC\over 2\sqrt {\epsilon-BC}}
$$


\begin{thebibliography}{9}

\bibitem{i} A.N.Leznov  THE METHOD OF SCALAR $L-A$ PAIR AND THE SOLITON SOLUTIONS OF THE PERIODIC TODA LATTICE {\it Proceedings of II INTERNATIONAL WORKSHOP "NONLINEAR AND TURBULENT PROCESSOS" Kiev 1983},{\it GORDON AND BREACH NEW-YORK 1984}(1437-1453)

A.N.Leznov {\it FUNCTIONAL ANAL. APPL. 18, (83-86), 1984}

A.N.Leznov {\it LETT.MATH.PHYS. v8, N5 (379-385) 1984} 

A.N.Leznov {\it LETT.MATH.PHYS. v8, N4 (353-358) 1984} 

\bibitem{7} A.N.Leznov  and A.V.Razumov {\it J.MATH.PHYS v35, (4067-4080), 1994}

\bibitem{8} A.N.Leznov  and A.V.Razumov {\it J.MATH.PHYS v35, (1738-1754), 1994}

\bibitem{10} A.N.Leznov  COMPLETELY INTEGRABLE SYSTEMS. {\it PREPRINT IHEP 92-112,
(1-66), 1992}

\bibitem{11} A.N.Leznov {\it Physics of elementary particals and atom nuclears}
N27, v.5, p 1161-1246 (1996)

\bibitem{LS} A.N.Leznov and A.S.Sorin {\it PHYS.LETT.B389:494-502,1996}.

\bibitem{LY} A. N. Leznov and E. A. Yusbashyan {\it LMP v35, p. 345-349, (1995)}

\bibitem{LYY} A. N. Leznov and E. A. Yusbashyan {\it Nucl.Phys.B 496,(3),643-653,(1997)}

\bibitem{FT} L.D.Fadeev and L.A.Takhtajan {\it Hamiltonian approach in soliton theory Moskow, Nauka (1985)}

\end{thebibliography}
\end{document}